\documentclass[
 reprint,
 superscriptaddress,
 amsmath,amssymb,
 aps,
 pra,
 tightenlines,
 twocolumn,
]{revtex4-2}

\usepackage{graphicx}
\usepackage{dcolumn}
\usepackage{bm}
\usepackage{color}
\usepackage{ulem}

\begin{document}

\title{Ground-state magnetic structures of topological kagome metals \textit{R}V$_6$Sn$_6$ (\textit{R} = Tb, Dy, Ho, Er)}

\author{Yishui Zhou}
\affiliation{J\"ulich Centre for Neutron Science (JCNS) at the Heinz Maier-Leibnitz Zentrum (MLZ), Forschungszentrum J\"ulich, Lichtenbergstrasse 1, D-85747 Garching, Germany}
\affiliation{Technical University of Munich (TUM), TUM School of Natural Sciences, Physics Department, D-85747 Garching, Germany}

\author{Min-Kai Lee}
\affiliation{Department of Physics, National Cheng Kung University (NCKU), Tainan 70101, Taiwan}

\author{Sabreen Hammouda}
\affiliation{J\"ulich Centre for Neutron Science (JCNS) at the Heinz Maier-Leibnitz Zentrum (MLZ), Forschungszentrum J\"ulich, Lichtenbergstrasse 1, D-85747 Garching, Germany}

\author{Sheetal Devi}
\affiliation{J\"ulich Centre for Neutron Science (JCNS) at the Heinz Maier-Leibnitz Zentrum (MLZ), Forschungszentrum J\"ulich, Lichtenbergstrasse 1, D-85747 Garching, Germany}

\author{Shin-Ichiro Yano}
\affiliation{National Synchrotron Radiation Research Center, Hsinchu 30077, Taiwan}

\author{Romain Sibille}
\affiliation{Laboratory for Neutron Scattering and Imaging, PSI Center for Neutron and Muon Sciences, Forschungsstrasse 111, 5232 Villigen, PSI, Switzerland}

\author{Oksana Zaharko}
\affiliation{Laboratory for Neutron Scattering and Imaging, PSI Center for Neutron and Muon Sciences, Forschungsstrasse 111, 5232 Villigen, PSI, Switzerland}

\author{Wolfgang Schmidt}
\affiliation{J\"ulich Centre for Neutron Science (JCNS) at ILL, Forschungszentrum J\"ulich, F-38000 Grenoble, France}

\author{Karin Schmalzl}
\affiliation{J\"ulich Centre for Neutron Science (JCNS) at ILL, Forschungszentrum J\"ulich, F-38000 Grenoble, France}

\author{Ketty Beauvois}
\affiliation{Universit\'e Grenoble Alpes, CEA, IRIG, MEM, MDN, F-38000 Grenoble, France}

\author{Eric Ressouche}
\affiliation{Universit\'e Grenoble Alpes, CEA, IRIG, MEM, MDN, F-38000 Grenoble, France}

\author{Po-Chun Chang}
\affiliation{J\"ulich Centre for Neutron Science (JCNS) at the Heinz Maier-Leibnitz Zentrum (MLZ), Forschungszentrum J\"ulich, Lichtenbergstrasse 1, D-85747 Garching, Germany}
\affiliation{Department of Physics, Tamkang University, Tamsui 251301, Taiwan}

\author{Chun-Hao Huang}
\affiliation{J\"ulich Centre for Neutron Science (JCNS) at the Heinz Maier-Leibnitz Zentrum (MLZ), Forschungszentrum J\"ulich, Lichtenbergstrasse 1, D-85747 Garching, Germany}

\author{Lieh-Jeng Chang}
\email{ljchang@ncku.edu.tw}
\affiliation{Department of Physics, National Cheng Kung University (NCKU), Tainan 70101, Taiwan}

\author{Thomas Br\"uckel}
\affiliation{J\"ulich Centre for Neutron Science JCNS and Peter Grünberg Institut PGI, JARA-FIT, Forschungszentrum J\"ulich, D-52425 J\"ulich, Germany}

\author{Yixi Su}
\email{y.su@fz-juelich.de}
\affiliation{J\"ulich Centre for Neutron Science (JCNS) at the Heinz Maier-Leibnitz Zentrum (MLZ), Forschungszentrum J\"ulich, Lichtenbergstrasse 1, D-85747 Garching, Germany}

\date{\today}

\begin{abstract} 

Magnetic kagome metals have attracted tremendous research interests recently, because they represent an ideal playground for exploring the fascinating interplay between their intrinsically inherited topologically non-trivial electron band structures, magnetism and electronic correlation effects, and the resultant novel electronic/magnetic states and emergent excitations. In this work, we report a comprehensive single-crystal neutron diffraction investigation of the ground-state magnetic structures of the recently discovered V-based topological kagome metals \textit{R}V$_6$Sn$_6$ (\textit{R} = Tb, Dy, Ho, Er). Furthermore, the sample synthesis details and our systematic studies of crystal structure, low-temperature magnetic and thermodynamic properties of these compounds via various in-house characterization techniques are also reported. Our single-crystal neutron diffraction measurements confirm that the long-range magnetic order forms below 4.3 K for $R$ = Tb, 3.0 K for $R$ = Dy, 2.4 K for $R$ = Ho, and 0.6 K for  $R$ = Er, respectively. The ground-state magnetic structures of the studied compounds are comprehensively determined via the magnetic crystallography approaches. It can be revealed that \textit{R}V$_6$Sn$_6$ (\textit{R} = Tb, Dy, Ho) have a collinear ferromagnetic order in the ground state, with the ordered magnetic moment aligned along the \textit{c} axis for \textit{R} = Tb, Ho, while approximately 20${^\circ}$ tilted off from the \textit{c} axis for \textit{R} = Dy. In contrast, ErV$_6$Sn$_6$ shows an A-type antiferromagnetic structure with a magnetic propagation vector \textbf{k} = (0, 0, 0.5), and with the ordered magnetic moment aligned in the \textit{ab} plane. The ordered magnetic moment are determined as 9.4(2) ${\mu_B}$, 6.6(2) ${\mu_B}$, 6.4(2) ${\mu_B}$, and 6.1(2) ${\mu_B}$ for \textit{R} = Tb, Dy, Ho, and Er, respectively. A comparison of the low-temperature magnetic structures for both the extensively investigated topological kagome metal series of \textit{R}V$_6$Sn$_6$ and \textit{R}Mn$_6$Sn$_6$ is given in details. This allows to gain new insights into the complex magnetic interactions, diverse single-ion magnetic anisotropies and spin dynamics in these compounds. The reported ground-state magnetic structures in \textit{R}V$_6$Sn$_6$ (\textit{R} = Tb, Dy, Ho, Er) can pave the way for further explorations of the possible interplay between magnetism and topologically non-trivial electron band structures in the magnetically ordered phase regime.

\end{abstract}

\maketitle

\section{Introduction}

Topological metals and semimetals possessing a two-dimensional kagome lattice are an ideal platform for the studies of quantum interactions between geometric effects of lattice, topologically non-trivial electron band structures, magnetism, and electronic correlations, owing to some extraordinary features in their electronic structures such as Dirac cones, flat bands, and van Hove singularities \cite{yin_topological_2022, ghimire_topology_2020, xu_quantum_2023, neupert_charge_2022}. For instance, Dirac cones are a defining feature in electronic structure that can lead to topologically protected states, flat bands can be a fertile ground for unconventional superconductivity, and van Hove singularities can lead to instabilities towards exotic electronic orders \cite{kang_dirac_2020, liu_orbital-selective_2020, li_dirac_2021}. Emergent correlated topological phases are widely found in kagome metals and semimetals, such as magnetic Weyl fermions in Mn$_3$Sn \cite{nakatsuji_large_2015, ikhlas_large_2017, kimata_magnetic_2019, tsai_electrical_2020, li_field-linear_2023} and Co$_3$Sn$_2$S$_2$ \cite{wang_large_2018, liu_magnetic_2019, morali_fermi-arc_2019, xing_localized_2020, howard_evidence_2021}, massive Dirac fermions in Fe$_3$Sn$_2$ \cite{kida_giant_2011, ye_massive_2018, lin_flatbands_2018, yin_giant_2018, li_magnetic-field_2019}, quantum-limited magnetic Chern phases in TbMn$_6$Sn$_6$ \cite{yin_quantum-limit_2020, xu_topological_2022, mielke_iii_low-temperature_2022, riberolles_orbital_2023-1}, charge-density waves (CDW) in FeGe \cite{teng_magnetism_2023} and ScV$_6$Sn$_6$ \cite{arachchige_charge_2022, hu_optical_2023, di_sante_flat_2023,tan_abundant_2023, pokharel_frustrated_2023, korshunov_softening_2023, guguchia_hidden_2023, subedi_order-by-disorder_2024, hu_phonon_2024}, and both CDW and superconductivity in \textit{A}V$_3$Sb$_5$ (\textit{A} = K, Rb, Cs) \cite{jiang_unconventional_2021, chen_roton_2021, yang_giant_2020, ortiz_new_2019, zhao_cascade_2021}.
 
Among them, the intermetallic \textit{R}T$_6$Sn$_6$ (\textit{R} = rare earths, \textit{T} = transition metals) compounds stand out for its unique crystal structure of HfFe$_6$Ge$_6$-type (space group No. 191, P6/mmm), that has a kagome bilayer structure formed by transition metal element and in-between a triangular layer of rare-earth element stacked along the \textit{c} axis in the unit cell. Recently, the \textit{R}Mn$_6$Sn$_6$ series has attracted a lot of attention not only for its complex magnetism due to rich magnetic interactions from the Mn-Mn, $R$-Mn, and $R$-$R$ couplings, but also for the intrinsic tunability of the magnetic properties by rare-earth elements that have distinct single-ion magnetic anisotropies \cite{venturini_magnetic_1991, venturini_magnetic_1993, venturini_incommensurate_1996, malaman_magnetic_1999, el_idrissi_magnetic_1991, kimura_high-field_2006, yin_quantum-limit_2020, ghimire_competing_2020, dally_chiral_2021, dhakal_anisotropically_2021, wang_field-induced_2021-1, li_dirac_2021, ma_anomalous_2021, ma_rare_2021, gao_anomalous_2021, zhang_exchange-biased_2022, riberolles_orbital_2023-1, riberolles_new_2024}. The \textit{R}Mn$_6$Sn$_6$ (\textit{R} = Gd, Tb, Dy, and Ho) compounds are all ferrimagnetic (FiM) but with different magnetic anisotropy, for instance, in-plane for \textit{R} = Gd, out-of-plane for \textit{R} = Tb, tilted off from the \textit{c} axis for \textit{R} = Dy, Ho, while other \textit{R}Mn$_6$Sn$_6$ compounds (\textit{R} = Sc, Y, Er, Tm, and Lu) have an incommensurate helical magnetic order over a large temperature range. Such a high diversity in magnetic anisotropy and magnetic structure allows the rare-earth engineering of various topological quantum phases \cite{ma_anomalous_2021}.

Similarly, the kagome metal \textit{R}V$_6$Sn$_6$ series was recently discovered and has also attracted strong interests due to rich quantum phenomena and high diversity in magnetic properties, in which non-magnetic V replaces Mn in the kagome layers, and the magnetic interaction is mainly contributed by the intralayer and interlayer $R$-$R$ couplings. For the non-magnetic \textit{R}V$_6$Sn$_6$ (\textit{R} = Sc, Y, Lu) compounds, the CDW was found in ScV$_6$Sn$_6$ at 92 K \cite{arachchige_charge_2022} and thus it has attracted a huge attention \cite{hu_optical_2023, di_sante_flat_2023,tan_abundant_2023, pokharel_frustrated_2023, korshunov_softening_2023, guguchia_hidden_2023, subedi_order-by-disorder_2024, hu_phonon_2024}, the electronic properties of the non-magnetic YV$_6$Sn$_6$ exhibit high mobility and multiband transport \cite{pokharel_electronic_2021, lee_anisotropic_2022}, but there is not so much study on LuV$_6$Sn$_6$ besides specific heat capacity \cite{lee_anisotropic_2022}. For the magnetic \textit{R}V$_6$Sn$_6$ (\textit{R} = Gd, Tb, Dy, Ho, Er, Tm) compounds, the magnetic transition temperatures are 4.9 K, 4.2 K, 3.0 K, and 2.4 K for \textit{R} = Gd, Tb, Dy, and Ho, respectively, but no magnetic order observed for \textit{R} = Er and Tm down to 1.8 K \cite{lee_anisotropic_2022, zhang_electronic_2022}. Recent angle-resolved photon-emission spectroscopy (ARPES) studies of GdV$_6$Sn$_6$ and HoV$_6$Sn$_6$ found indeed Dirac cones, saddle points, and flat bands in electronic structure \cite{peng_realizing_2021}, and proved the presence of the topologically non-trivial states in GdV$_6$Sn$_6$ \cite{hu_tunable_2022}. Furthermore, a finite spin Berry curvature in \textit{R}V$_6$Sn$_6$ (\textit{R} = Tb, Ho, Sc) has been unveiled \cite{di_sante_flat_2023}. Besides, an incommensurate magnetic order with the magnetic moments oriented in the \textit{ab} plane was found in GdV$_6$Sn$_6$ via resonant X-ray diffraction \cite{porter_incommensurate_2023}, while the magnetism in TbV$_6$Sn$_6$ was found highly anisotropic with easy axis along the \textit{c} axis \cite{rosenberg_uniaxial_2022, pokharel_highly_2022}. In addition, partial crystallographic disorder and a large easy-plane magnetic anisotropy were found in SmV$_6$Sn$_6$ \cite{huang_anisotropic_2023-1}, and YbV$_6$Sn$_6$ was discovered as a heavy-fermion compound hosting a triangular Kondo lattice \cite{guo_triangular_2023}. A study of the magnetic and magnetotransport properties on DyV$_6$Sn$_6$ and HoV$_6$Sn$_6$ found that they have ferromagnetic interaction along the \textit{c} axis and antiferromagnetic interaction within the \textit{ab} plane and are multi-band systems \cite{zeng_magnetic_2024}. Despite extensive recent investigations on the magnetic properties of the kagome metals \textit{R}V$_6$Sn$_6$ (\textit{R} = Tb, Dy, Ho, Er) \cite{pokharel_highly_2022,lee_anisotropic_2022, zhang_electronic_2022,rosenberg_uniaxial_2022,zeng_magnetic_2024}, their ground-state magnetic structures, which are essential for the understanding of the intertwined topological and magnetic properties in these compounds, have not been determined so far.

In this work, we report a comprehensive single-crystal neutron diffraction study of the  low-temperature magnetic properties of the recently discovered V-based topological kagome metals of \textit{R}V$_6$Sn$_6$ (\textit{R} = Tb, Dy, Ho, Er). Our single-crystal neutron diffraction measurements confirm the occurrence of the long-range magnetic order at 4.3 K for $R$ = Tb, 3.0 K for $R$ = Dy, 2.4 K for $R$ = Ho, and 0.6 K for $R$ = Er, respectively. Based on the magnetic crystallography approaches, the ground-state magnetic structures of the studied compounds are comprehensively determined. The \textit{R}V$_6$Sn$_6$ (\textit{R} = Tb, Dy, Ho) compounds have a collinear ferromagnetic order below the respective magnetic phase transition temperature, with the ordered magnetic moment aligned along the \textit{c} axis for \textit{R} = Tb, Ho, while approximately 20${^\circ}$ tilted off from the \textit{c} axis for \textit{R} = Dy. ErV$_6$Sn$_6$ shows an A-type antiferromagnetic structure with a magnetic propagation vector \textbf{k} = (0, 0, 0.5), and with the ordered moment aligned in the \textit{ab} plane. The ordered magnetic moment are determined as 9.4(2) ${\mu_B}$, 6.6(2) ${\mu_B}$, 6.4(2) ${\mu_B}$, and 6.1(2) ${\mu_B}$ for \textit{R} = Tb, Dy, Ho, and Er, respectively. In addition, we also report the sample synthesis via the flux-method crystal growth, and the systematic investigations of crystal structures, magnetic properties and specific heat capacity (down to 50 mK) of the high-quality single-crystal samples of this kagome metal series via a wide range of in-house characterization techniques. A comparison of the  low-temperature magnetic structures in both \textit{R}V$_6$Sn$_6$ and \textit{R}Mn$_6$Sn$_6$ kagome metal series is given. This allows to shed light on the complex magnetic interactions, diverse single-ion magnetic anisotropies and spin dynamics in these compounds. Our study would motivate further investigations on the possible interplay between magnetism and topologically non-trivial electron band structures in the magnetically ordered phase regime in this fascinating topological kagome metal series of \textit{R}V$_6$Sn$_6$.

\begin{figure}[htpb]
\centering
\includegraphics[width=8cm]{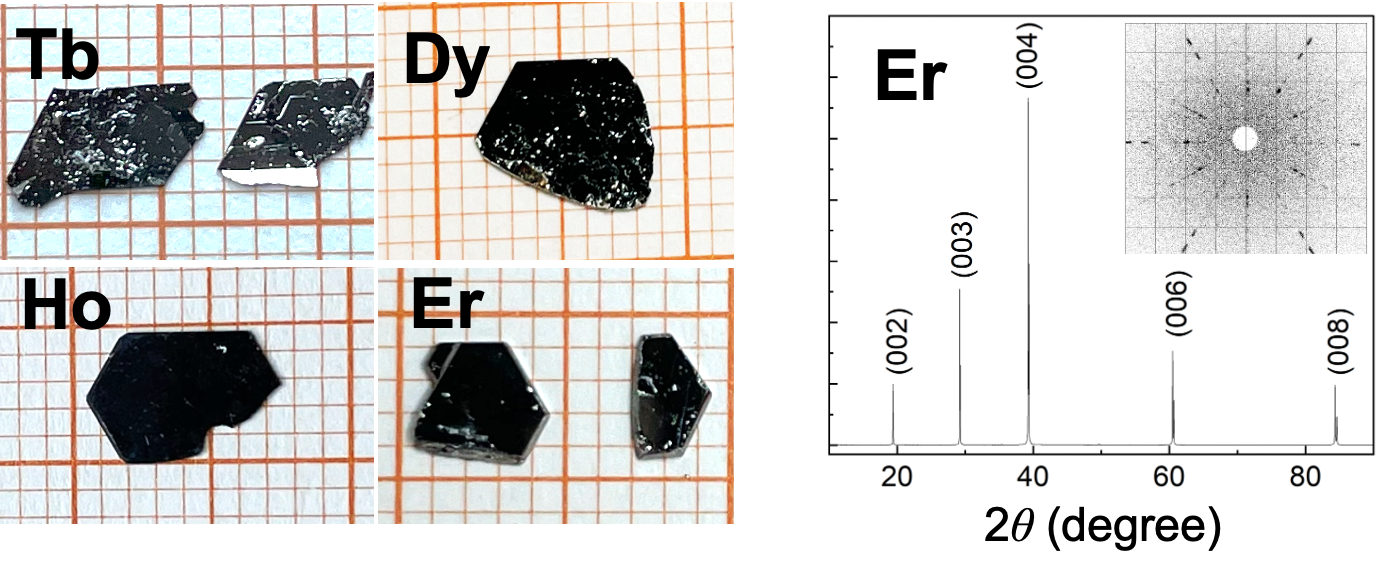}
\hspace{0cm}
\caption{\label{fig:Figs_crystals}
Single crystals of \textit{R}V$_6$Sn$_6$ (\textit{R} = Tb, Dy, Ho, Er) via self-flux method and X-ray diffraction pattern of ErV$_6$Sn$_6$ single crystal in the \textbf{c} direction, the inset shows the Laue X-ray pattern in the same direction.
}
\end{figure}

\section{Experimental details}

Single crystals of \textit{R}V$_6$Sn$_6$ were grown via the self-flux method. Rare-earth element pieces, V granular, and Sn shots were loaded in a crucible in the glove box with the molar ratio of $R: V: Sn = 1: 2: 40$, which was then sealed in a quartz tube under an argon atmosphere with the pressure of 200 mbar. The tube was heated to 1200 $^{\circ}$C over 12 hours and then dwelt for 10 hours. Afterwards, it was slowly cooled down to 800 $^{\circ}$C with a cooling rate of 1 $^{\circ}$C/h followed by centrifuging to separate crystals from the flux, and then shiny plate-shape crystals were obtained (see Fig. \ref{fig:Figs_crystals}).

In-house single-crystal X-ray diffraction (XRD) was measured using a Rigaku XtaLAB Synergy-S diffractometer with MoK$_{\alpha}$. The CrysAlis$^{Pro}$ software is used to search for a proper unit cell, and after indexing, to integrate the intensity of the Bragg reflections over different image frames. The absorption correction is done using the indexed crystal facets. The Jana2006 program is used for the crystal structure refinement and the charge flipping approach for solving crystal structures \cite{petricek_crystallographic_2014}.

A few selected single crystals were used for measuring the magnetic properties by SQUID from Quantum Design. Magnetic susceptibility ($M/H-T$) was measured with sweeping temperature from 300 to 2 K under a magnetic field of 1 T. The isothermal magnetization ($M-H$) curves were measured from 0 to 5 T at different temperatures.

The molar heat capacity at constant pressure (C$_p$) of the \textit{R}V$_6$Sn$_6$ single crystals was measured using a PPMS device from Quantum Design. After the addenda measurement, the samples with suitable mass and size were chosen to optimize the thermal coupling between the sample and the puck. Specific heat was measured in two temperature ranges, from 50 mK to 3 K and from 2 K to 100 K, and then the data were combined accordingly. Note that only the low-temperature part for DyV$_6$Sn$_6$ and only the high-temperature part for LuV$_6$Sn$_6$ were measured.

Comprehensive single-crystal neutron diffraction experiments were carried out with a series of neutron instruments at various neutron facilities. For TbV$_6$Sn$6$, a 3 mg high-quality single crystal was chosen and mounted on an aluminium pin with GE-vanish. The neutron diffraction experiment was conducted at the thermal neutron single-crystal diffractometer Zebra at SINQ, PSI. A neutron beam with wavelength of 1.383 \AA\ was generated using a Ge(311) monochromator. An open HUBER cradle combined with a dedicated 1.6-300 K cryostat (JT-CCR) was used. Initially, an area detector was utilized to locate the peaks and subsequently switched to a single detector configuration for data collection. Neutron diffraction measurements on the HoV$_6$Sn$_6$ and ErV$_6$Sn$_6$ single crystals were initially performed using the cold neutron triple-axis spectrometer (TAS) Sika at ANSTO. The samples, weighing 33.2 mg and 42.6 mg, respectively, were mounted in the ($H$, $H$, $L$) horizontal scattering plane. All elastic scans were performed with $\textbf{k}_i=\textbf{k}_f=2.662$ \AA$^{-1}$. A PG filter and collimators with a sequence of $0'-0'-60'-60'$ were employed in the experiment. A dilution refrigerator (DF-1) was used for reaching the base temperature of 50 mK for ErV$_6$Sn$_6$. Due to the limited reflections accessible with a cold TAS instrument, while the magnetic propagation vector and the phase transition temperature for both compounds were determined, their magnetic structures could not be solved based on a small number of reflections in the horizontal scattering plane. Further single-crystal neutron diffraction experiments on DyV$_6$Sn$_6$, HoV$_6$Sn$_6$ and ErV$_6$Sn$_6$, with a sample weight at 36.1, 40.9 and 30.7 mg, respectively, were carried out on the thermal neutron diffractometer D23 at ILL. All measured single-crystal samples at D23 were also mounted in the ($H$, $H$, $L$) horizontal scattering plane, nevertheless, the lifting-counter setup at D23 allows the measurement of out-of plane reflections. A wavelength of 1.283 \AA\ was employed for data collection. A standard ILL 1.5-300 K orange cryostat combined with a dilution insert was used for ErV$_6$Sn$_6$. All neutron diffraction experiments were carried out under zero magnetic field, and the refinement of magnetic structure as well as neutron absorption correction have been done with the help of Mag2Pol \cite{qureshi_mag2pol_2019}. The momentum transfer \textbf{Q} = ($H$, $K$, $L$) is defined in hexagonal reciprocal lattice units (r.l.u.) based on the lattice constants given in Tab. \ref{tab:tab1} and Tab. \ref{tab:neutron}, where $H$, $K$, and $L$ are miller indices. Both \textbf{Q} and sample-rotation (i.e., Omega) scans were measured for various reflections.

\begin{figure}[htpb]
\centering
\includegraphics[width=8cm]{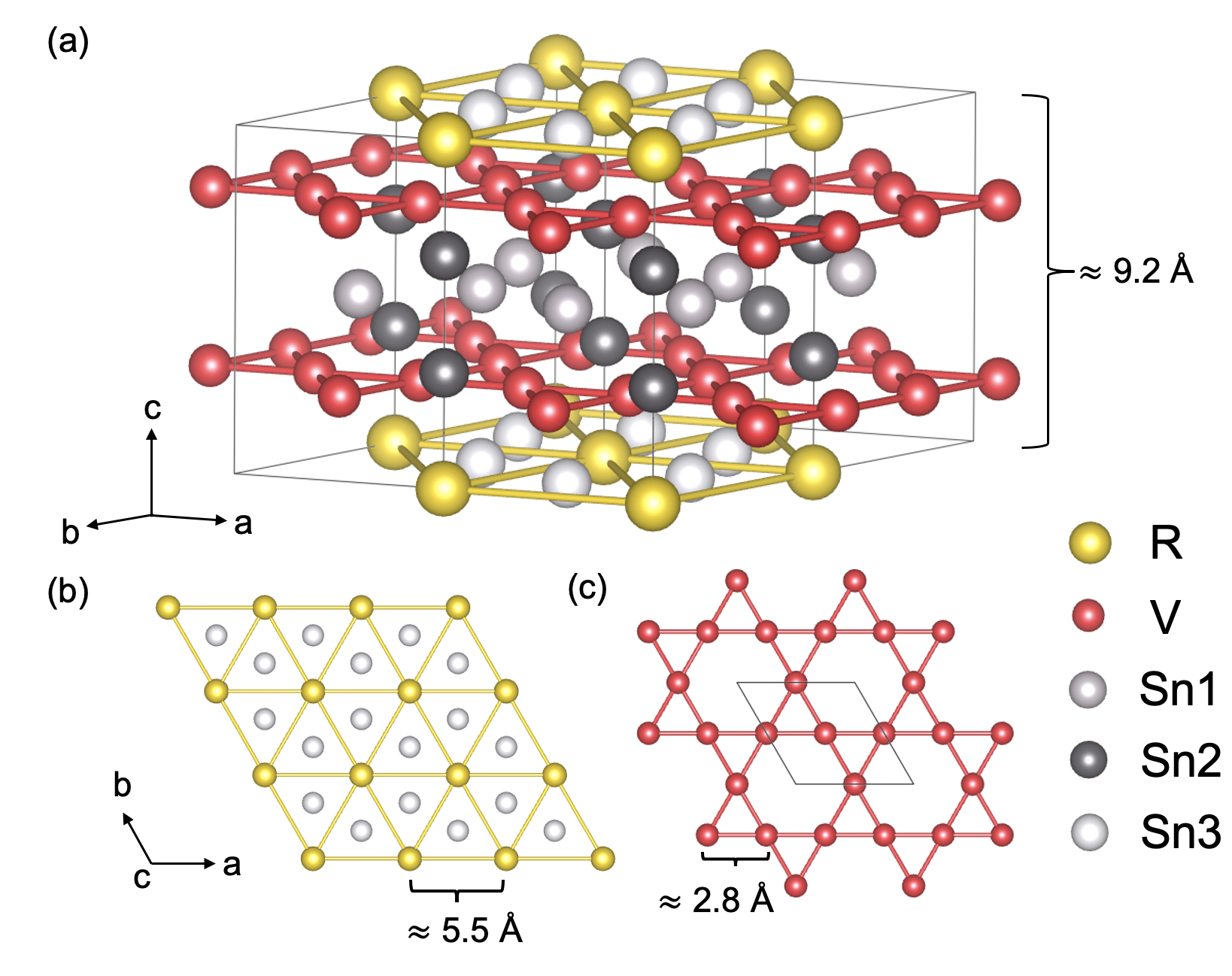}
\hspace{0cm}
\caption{\label{fig:Figs_structure}
Crystal structure of \textit{R}V$_6$Sn$_6$.
(a) Hexagonal structure formed by \textit{R}Sn, V, and Sn layers, with the order of [-\textit{R}Sn2-V-Sn3-Sn1-Sn3-V-], and the distance between two adjacent \textit{R} layers is about 9.2 \AA. 
(b) Triangular structure of \textit{R} sublattice from the top view, with Sn2 in the middle of each triangular, and the distance between the nearest \textit{R} ions in the same layer is about 5.5 \AA.
(c) Kagome structure of V sublattice from the top view, the distance between the nearest V ions is about 2.8 \AA.
}
\end{figure}

\section{In-house characterization results}

\subsection{X-Ray Diffraction and Crystalline Structures}

The crystal structure of \textit{R}V$_6$Sn$_6$ (\textit{R} = Tb, Dy, Ho, Er, Lu) is determined through the refinement of single-crystal XRD data collected at room temperature. As depicted in Fig. \ref{fig:Figs_structure}(a), all compounds exhibit a HfFe$_6$Ge$_6$-type hexagonal structure with the space group $P6/mmm$ (No. 191). In this structure, \textit{R}(1a) and V(6i) ions occupy a single Wyckoff position, while Sn ions are found in three distinct Wyckoff positions, namely, Sn1(2d), Sn2(2e), and Sn3(2c). Along the \textit{c} axis, the layers are arranged in the sequence of [-\textit{R}Sn3-V-Sn2-Sn1-Sn2-V-]. \textit{R} ions form a triangular layer with Sn3 situated at the center of each triangle (see Fig. \ref{fig:Figs_structure}(b)), with an approximate inter-ionic distance of 5.5 \AA\ between the nearest \textit{R} ions. Within a single layer, V ions adopt a perfect kagome structure (see Fig. \ref{fig:Figs_structure}(c)), with the nearest V-V distance at around 2.8 \AA. The unit cell contains two V kagome layers, and the distance between adjacent \textit{R} layers, also referred to as the \textit{c} lattice parameter, is approximately 9.2 \AA. Further details regarding the lattice parameters can be found in Tab. \ref{tab:tab1}.

\begin{figure*}[htpb]
\centering
\includegraphics[width=16cm]{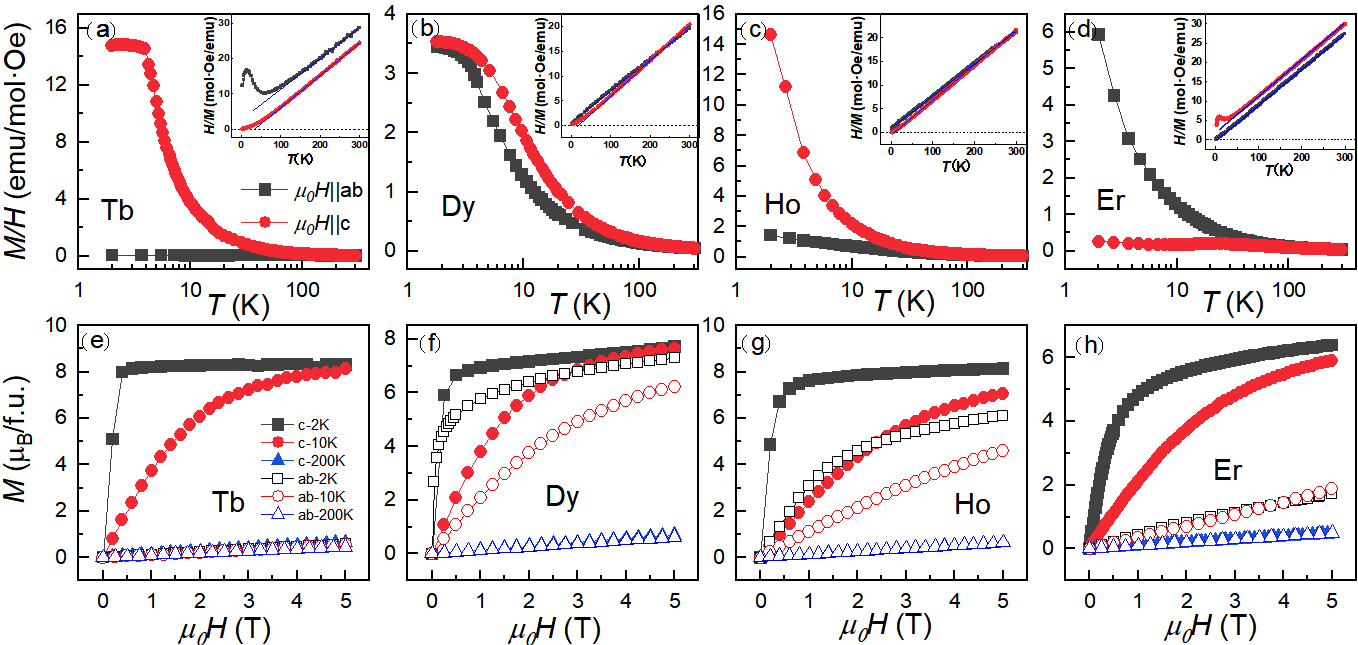}
\hspace{0cm}
\caption{\label{fig:Figs_magnetic_properties}
Magnetic properties of \textit{R}V$_6$Sn$_6$ (\textit{R} = Tb, Dy, Ho, Er), measured with the magnetic field applied within the \textit{ab} plane ($\mu_0H||\textit{ab}$) and along the \textit{c} axis ($\mu_0H||\textit{c}$).
(a$-$d) shows the susceptibilities under the magnetic field of 1 T measured during cooling, and the insets are the reciprocal of susceptibilities and Curie-Weiss fitting. (e$-$h) shows the corresponding isothermal magnetization properties at different temperatures.
}	
\end{figure*}

	\begin{table*}[htpb]
		\caption[\textit{R}V$_6$Sn$_6$ refinement results]{\label{tab:tab1}
    Crystal parameters and physical properties of \textit{R}V$_6$Sn$_6$ (\textit{R}$=$ Tb$-$Er, Lu). 
    $a$, $c$, and $V$ are refined crystal parameters of \textit{R}V$_6$Sn$_6$ of X-ray single crystal diffraction at room temperature. 
    $\theta_{CW}^{ab}$ and $\theta_{CW}^{c}$ are the fitted Curie temperatures obtained from the Curie-Weiss fitting on susceptibility with the magnetic field applied within the \textit{ab} plane and along the \textit{c} axis. 
    $\mu_{eff}^{ab}$ and $\mu_{eff}^{c}$ are the fitted effective magnetic moments for \textit{R}$^{3+}$ ion according to the Curie constant, and the $\mu_{eff}^{free}(\mu_B)$ is the effective moment of free \textit{R}$^{3+}$ ion. 
    $R_{ani}$ represents the anisotropy ratio, obtained from the susceptibility in the easy-axis direction over in the hard-axis direction at 2 K under the magnetic field of 5 T.  
    $T_{C/N}(K)$ is the transition temperature obtained by heat capacity measurements under zero field (Note that the $T_C$ value of TbV$_6$Sn$_6$ is taken from the DC magnetization measurement).
   }
   
    \begin{ruledtabular}
    	  \begin{tabular}{ccccccccccc}   
      & Parameter & TbV$_6$Sn$_6$ & DyV$_6$Sn$_6$ & HoV$_6$Sn$_6$ & ErV$_6$Sn$_6$ & LuV$_6$Sn$_6$ \\
      & $a(\AA)$ & 5.5214(2) & 5.5161(2) & 5.5157(4) & 5.5113(2) & 5.4976(2)\\
      & $c(\AA)$ & 9.1791(2) & 9.1872(6) & 9.1774(2) & 9.1752(2) & 9.1846(2)\\
      & $V(\AA^3)$ & 242.34(1) & 242.09(1) & 241.80(1) & 241.35(1)& 240.40(1)\\
      & $\theta_{CW}^{ab}(K)$ & -49.29 & -11.88 & -5.21 & 0.76 & -\\
      & $\theta_{CW}^{c}(K)$ & 33.92 & 6.48 & 9.29 & -25.08 & -\\
      & $\mu_{eff}^{ab}(\mu_B)$ & 10.28 & 11.19 & 10.41 & 9.41 & -\\
      & $\mu_{eff}^{c}(\mu_B)$ & 9.29 & 11.09 & 10.41 & 9.42 & -\\
      & $\mu_{eff}^{free}(\mu_B)$ & 9.72 & 10.65 & 10.61 & 9.58 & -\\
      & Easy axis/plane & \textit{c}-axis & \textit{c}-axis & \textit{c}-axis & \textit{ab}-plane  & - \\
      & $R_{ani}$ & 181 & 1.2 & 10 & 23  & - \\
      & $T_{C/N}(K)$ & 4.0 & 2.9 & 2.3 & 0.5 & -\\
          \end{tabular}
       \end{ruledtabular}
   \end{table*}

\subsection{DC Magnetization}

The results from the direct current (DC) magnetization measurements of \textit{R}V$_6$Sn$_6$ (\textit{R} = Tb, Dy, Ho, Er) single crystals are shown in Fig. \ref{fig:Figs_magnetic_properties}, with each compound measured with the magnetic field applied within the \textit{ab} plane and along the \textit{c} axis. Fig. \ref{fig:Figs_magnetic_properties}(a-d) shows the susceptibilities under the magnetic field of 1 T and measured during cooling from 300 K to 2 K. These results demonstrate that the compounds display strong magnetic anisotropy. Specifically, TbV$_6$Sn$_6$ and HoV$_6$Sn$_6$ exhibit an easy axis along the \textit{c}-axis direction, while ErV$_6$Sn$_6$ exhibits an easy direction within the \textit{ab} plane. In comparison, the anisotropy in DyV$_6$Sn$_6$ is relatively weak, with the susceptibility along the \textbf{c} direction slightly larger than within the \textit{ab} plane. The insets in Fig. \ref{fig:Figs_magnetic_properties}(a-d) present reciprocal susceptibilities and Curie-Weiss fittings (indicated by blue lines) for each compound. Curie-Weiss fittings were performed in the temperature range of 150$-$300 K, using the formula $\chi(T)= \chi_0+C/(T-\theta_{CW})$. The values of the $\chi_0$ for these fittings are in the range of $10^{-4}-10^{-3}$ emu/mol Oe. The resulting Curie-Weiss temperatures for both in-plane ($\theta^{ab}_{CW}$) and out-of-plane ($\theta^{c}_{CW}$) directions are shown in Tab. \ref{tab:tab1}. The negative $\theta^{ab}_{CW}$ and positive $\theta^{c}_{CW}$ for \textit{R} = Tb, Dy, and Ho imply antiferromagnetic (AFM) interaction within the \textit{ab} plane and ferromagnetic (FM) interaction along the \textit{c} axis, while the situation for ErV$_6$Sn$6$ is on the contrary. The calculated effective magnetic moments, derived from the Curie-Weiss fitting, are also listed in Tab. \ref{tab:tab1} as $\mu_{eff}^{ab}$/$\mu_{eff}^{c}$, with values of 10.28/9.29 $\mu_B$, 11.19/11.09 $\mu_B$, 10.41/10.41 $\mu_B$, and 9.41/9.42 $\mu_B$ for Tb, Dy, Ho, and Er compounds, respectively. These values closely resemble those of the free $R^{3+}$ ions, indicating that the trivalent valence of rare earth elements and the nonmagnetic nature of V ions. Our susceptibility properties and the Curie-Weiss fitting results are consistent with the previous studies \cite{lee_anisotropic_2022, zhang_electronic_2022, zeng_magnetic_2024}.
 
Fig. \ref{fig:Figs_magnetic_properties}(e-h) presents the magnetization of these compounds at different temperatures, in which the anisotropy shows evidently, particularly at low temperatures. To assess the level of anisotropy, an anisotropic ratio $R_{ani}$ was defined as the ratio of magnetization in the easy direction to that in the hard direction at 2 K and under a 5 T magnetic field where the spin moments are more or less saturated. The values of $R_{ani}$ for these compounds are listed in Table I. TbV$_6$Sn$_6$ exhibiting extreme anisotropy along the \textit{c} axis, characterized by an $R_{ani}$ value of 181, suggests an Ising transition at low temperature which is consistent with former studies \cite{rosenberg_uniaxial_2022, pokharel_highly_2022}, while a very weak anisotropy along the \textit{c} axis of $R_{ani}$ value 1.2 for DyV$_6$Sn$_6$. The anisotropy for HoV$_6$Sn$_6$ is also along the \textit{c} axis and stronger than DyV$_6$Sn$_6$ but smaller than TbV$_6$Sn$_6$, with the anisotropy ratio of 10. ErV$_6$Sn$_6$ displays an easy direction within the \textit{ab} plane, and a relatively strong anisotropy with the $R_{ani}$ values of 23.

\begin{figure}[htpb]
\centering
\includegraphics[width=7cm]{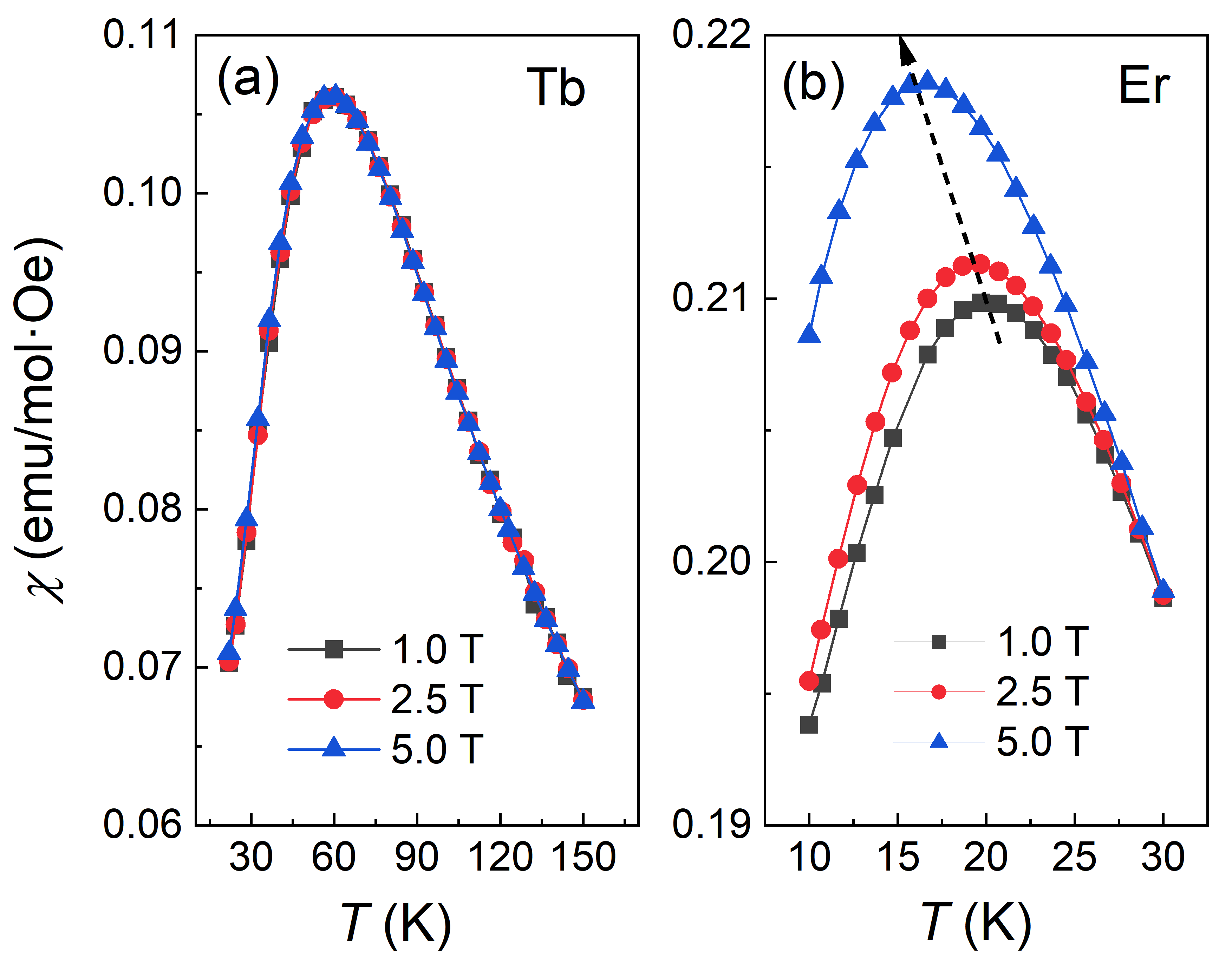}
\hspace{0.5cm}
\caption{\label{fig:Figs_broad_peak}
Broad peaks of susceptibility properties under different magnetic fields applied perpendicular to the easy-axis direction for (a) TbV$_6$Sn$_6$ and (b) ErV$_6$Sn$_6$.
}
\end{figure}

In addition, a broad feature is observed in the susceptibilities of TbV$_6$Sn$_6$ and ErV$_6$Sn$_6$ compounds, as shown in Fig. \ref{fig:Figs_broad_peak}, specifically along the direction perpendicular to the easy axis. Under different magnetic fields, the susceptibilities of TbV$_6$Sn$_6$ (see Fig. \ref{fig:Figs_broad_peak}(a)) compounds remain unchanged up to 5 T. However, in the case of ErV$_6$Sn$_6$ (see Fig. \ref{fig:Figs_broad_peak}(b)) compound, the broad peak shifts slightly to lower temperatures and exhibits higher susceptibility when a larger magnetic field is applied. A similar behavior was also observed in TbV$_6$Sn$_6$ by Pokharel $et$ $al.$ \cite{pokharel_highly_2022}. This broad-peak feature is likely due to the gradual development of single-ion magnetic anisotropy at low temperatures.

\begin{figure}[htpb]
\begin{center}
\hspace{1cm}\includegraphics[width=8cm]{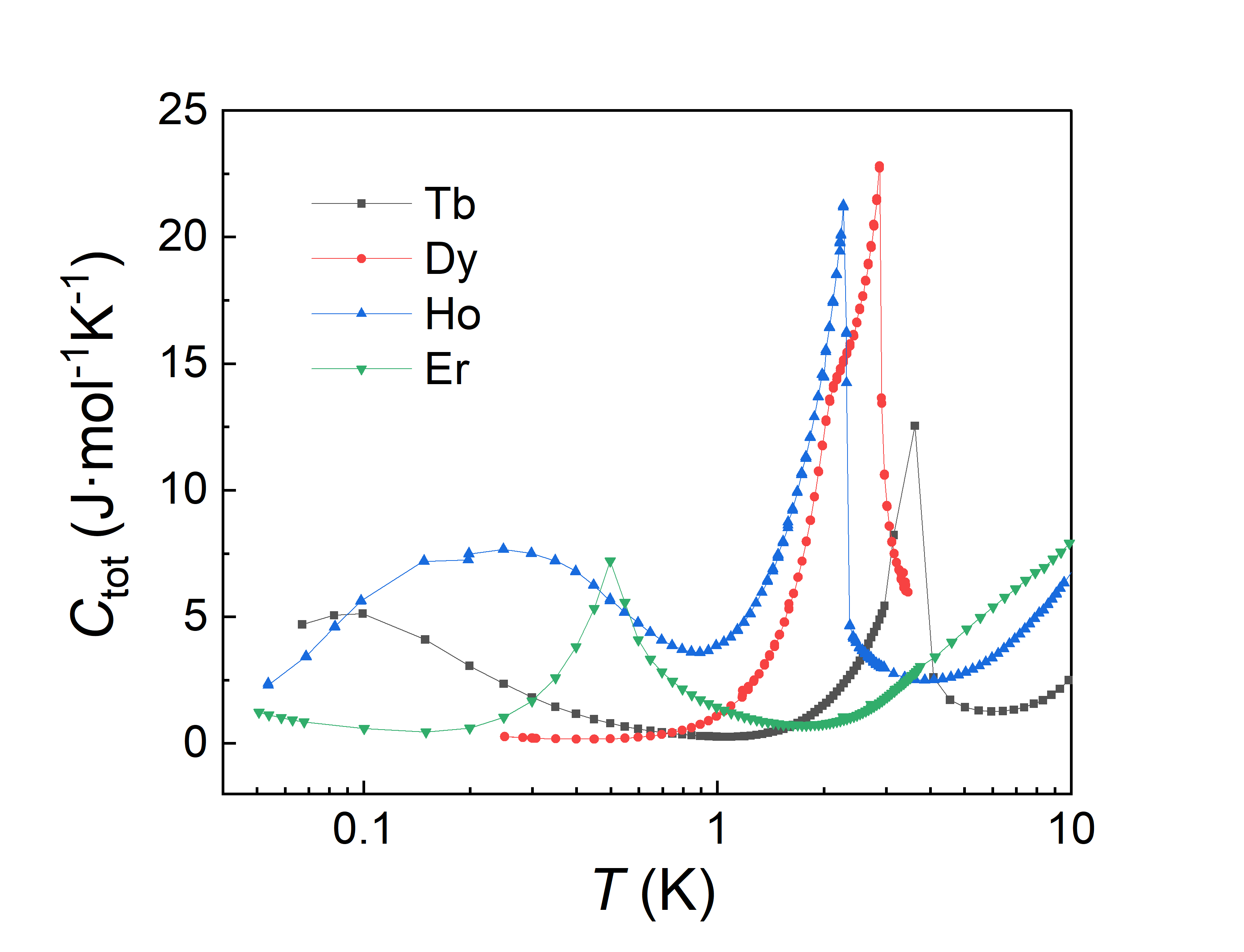}
\vspace{-5pt}
\caption{\label{fig:Figs_heat_capacity}
Heat capacity of \textit{R}V$_{6}$Sn$_{6}$ (\textit{R} = Tb, Dy, Ho, Er) compounds measured at zero magnetic field down to 50 mK.}
		\end{center}
	\end{figure}

\subsection{Heat Capacity}
	  
Fig. \ref{fig:Figs_heat_capacity} shows the temperature dependence of the specific heat for the compounds \textit{R}V$_6$Sn$_6$ (\textit{R} = Tb, Dy, Ho, and Er) down to 50 mK. In agreement with the magnetization studies, a $\lambda$-shape sharp heat capacity anomaly is observed at low temperatures for all the compounds, providing further support for the presence of long-range magnetic order at their ground states. The transition temperatures for Tb, Dy, Ho, and Er compounds are identified at $\sim$3.6 K, 2.9 K, 2.3 K, and 0.5 K, respectively. Our results are consistent with the recent studies on this series \cite{lee_anisotropic_2022, zhang_electronic_2022}, but it should be noted that the previous results are limited to 1.8 K. Besides, there is only one sharp peak for each compound down to 50 mK, and the additional broad or upturn feature below the long-range magnetic ordering temperature can be observed at very low temperatures in TbV$_{6}$Sn$_{6}$, HoV$_{6}$Sn$_{6}$, and ErV$_{6}$Sn$_{6}$, which is attributed to the nuclear Schottky heat-capacity contribution from the respective $R^{3+}$ ions. The nuclear Schottky anomaly is resulted from the splitting of the energy levels due to the hyperfine interaction between the electronic and nuclear spins of the $R^{3+}$ ions. Such hyperfine interactions can be rather large in some rare-earth elements such as Ho and Tb, thus leading to prominent nuclear Schottky anomaly that has widely been observed in various rare-earth based magnetic materials \cite{PhysRevLett.87.047205,Kumar_2016,PhysRevLett.94.246402,PhysRevX.9.031005}. A detailed analysis of the low-temperature heat capacity data of the studied compounds will be discussed in a separate work of ours together with non-magnetic LuV$_{6}$Sn$_{6}$.

	\begin{table*}[hbtp]
		\caption[RV$_6$Sn$_6$ neutron diffraction results]{\label{tab:neutron}
    Single-crystal neutron diffraction results for \textit{R}V$_6$Sn$_6$ (\textit{R} = Tb, Dy, Ho, Er). 
    $T_{C/N}(K)$ is the magnetic transition temperature obtained from the temperature dependent neutron diffraction measurements under zero field. The lattice parameters $a$ and $c$ are obtained based on the single-crystal neutron diffraction data taken below $T_{C/N}$ at the base temperature of 1.6 K, 1.8 K, 1.8 K, and 50 mK for $R$ = Tb, Dy, Ho, and Er, respectively. V(6$i$) and Sn2(2$e$) represent the refinable Wyckoff positions for the crystal structure  with space group $P6/mmm$. The refinements of the atomic positions of V(6$i$) and Sn2(2$e$) are performed based on the single-crystal neutron diffraction data taken both at HT (i.e., at 5.5, 5.0, 5.0, and 0.7 K for  Tb, Dy, Ho, and Er compounds, respectively), and at LT (i.e., at 1.6, 1.8, 1.8, and 0.05 K for Tb, Dy, Ho, and Er compounds, respectively).
    \textbf{k} is the magnetic propagation vector. 
    $\textbf{M}_{[1,0,0]}$, $\textbf{M}_{[1,2,0]}$, and $\textbf{M}_{[0,0,1]}$ are the projection of the magnetic moment along [1, 0, 0], [1, 2, 0], and [0, 0, 1] determined by the refinement, and the $\mu_{eff}^{free}(\mu_B)$ is the effective moment of free $R^{3+}$ ion. 
    The refinement parameter $R_F(Int)$ represents the average discrepancy between the observed and calculated values of the integrated intensity of the reflections, serving as a measure of the agreement between the experimental data and the theoretical model.
    }
    \begin{ruledtabular}
    	  \begin{tabular}{ccccccc}     
      & Parameter & TbV$_6$Sn$_6$ & DyV$_6$Sn$_6$ & HoV$_6$Sn$_6$ & ErV$_6$Sn$_6$ & \\
       \hline
      & $T_{C/N}(K)$ & 4.3 & 3.0 & 2.4 & 0.6 \\
      & $a(\AA)$ & 5.530(9) & 5.510(7) & 5.506(6) & 5.50(1) \\
      & $c(\AA)$ & 9.212(6) & 9.167(6) & 9.158(9) & 9.181(8) \\
      & V(6$i$) \;(HT) & 0.2562(4) & 0.2460(9) & 0.2553(9) & 0.2655(7) \\
      & V(6$i$) \;(LT) & 0.2488(4) & 0.2496(9) & 0.2517(7) & 0.2665(4) \\
      & Sn2(2$e$) \;(HT) & 0.3343(5) & 0.3339(5) & 0.3311(5) & 0.3321(8) \\
      & Sn2(2$e$) \;(LT) & 0.3346(5) & 0.3338(6) & 0.3298(6) & 0.3315(8) \\
      & $\textbf{k}$ & (0, 0, 0) & (0, 0, 0) & (0, 0, 0) & (0, 0, 0.5) \\
      & $|\textbf{M}_{[1,0,0]}|(\mu_B)$ & 0 & 2.3(2) & 0 & 6.1(3) \\
      & $|\textbf{M}_{[1,2,0]}|(\mu_B)$ & 0 & 0 & 0 & 0 \\
      & $|\textbf{M}_{[0,0,1]}|(\mu_B)$ & 9.4(2) & 6.2(1) & 6.4(2) & 0 \\
      & $\mu_{eff}^{free}(\mu_B)$ & 9.72 & 10.65 & 10.61 & 9.58 \\
      & $R_F(Int)$ & 1.96 & 5.53 & 3.84 & 3.09 \\
      \end{tabular}
    \end{ruledtabular}
     \end{table*}

\section{Magnetic structure determination via single-crystal neutron diffraction}

Having evidenced the presence of long-range magnetic order in their respective ground states, we turn to single-crystal neutron diffraction, a microscopic probe, to determine the magnetic structures of these four compounds. The first step is to identify the magnetic propagation vector of the ordered magnetic moment below its transition temperature. When the reflections with both structural and magnetic information are measured, there are several approaches to solve magnetic structure via refinement. If the magnetic propagation vector is commensurate, one can solve the magnetic structure employing the irreducible representations (Irreps) or magnetic space group (MSG). When the propagation vector is incommensurate, the approach using magnetic super-space group (MSSG) needs to be introduced.

In this work, all the studied \textit{R}V$_6$Sn$_6$ (\textit{R} = Tb, Dy, Ho, Er) compounds have a commensurate propagation vector, the Irreps are calculated with the space group P6/mmm by SARAh \cite{wills_new_2000} to obtain the allowed basis vectors. One difficulty in this series is that the V is almost invisible because of its rather small nuclear coherent scattering length for neutrons, so the atomic positions of both V(6$i$) and Sn2(2$e$) that are obtained from the structural refinement based on the XRD data taken at room temperature are used as the initial input parameters for the refinement of the neutron diffraction data. The refinements of the atomic positions of V(6$i$) and Sn2(2$e$) are performed based on the single-crystal neutron diffraction data taken at HT (i.e., slightly above $T_{C/N}$), and at LT (i.e., below $T_{C/N}$ at the base temperatures). No appreciable changes of the atomic positions of both V(6$i$) and Sn2(2$e$) can be seen (see Tab. \ref{tab:neutron}). Furthermore, the Omega scans, that are carefully performed for all samples at both HT and LT, do not show any changes in their peak position and peak width (see the following subsections). Therefore, we conclude that a concomitant structure phase transition at $T_{C/N}$ is unlikely, at least, within the detection limits of our neutron diffraction experiments. Since the absorption cross-sections of Dy and Er are about 994 and 159 barn for thermal neutrons, respectively, which is notably high when compared to other rare-rare elements, thus neutron absorption correction is needed. One can deal with the absorption correction via Mag2Pol \cite{qureshi_mag2pol_2019, zhu_magnetic_2020}, in which a 3-dimensional crystal model can be created based on the shape and size of the real sample on which the neutron diffraction is performed, then the intensity of the measured reflections can be corrected according to the distance that the neutron beam travels inside the sample. All the single-crystal diffraction results for \textit{R}V$_6$Sn$_6$ (\textit{R} = Tb, Dy, Ho, Er) are listed in Tab. \ref{tab:neutron}. It is worth noting that the lattice parameters of TbV$_6$Sn$_6$ at 1.6 K obtained via neutron diffraction (see Tab. \ref{tab:neutron}) are slightly larger than that obtained at room temperature via XRD (see Tab. \ref{tab:tab1}). This is rather unusual and potentially interesting, further experiments, such as low-temperature XRD measurements, or low-temperature strain measurements on single-crystal samples using strain gauges and capacitance bridges, would be very helpful to clarify this.

\subsection{Magnetic Structure of TbV$_6$Sn$_6$}

\begin{figure}[htpb]
\centering
\includegraphics[width=8.45cm]{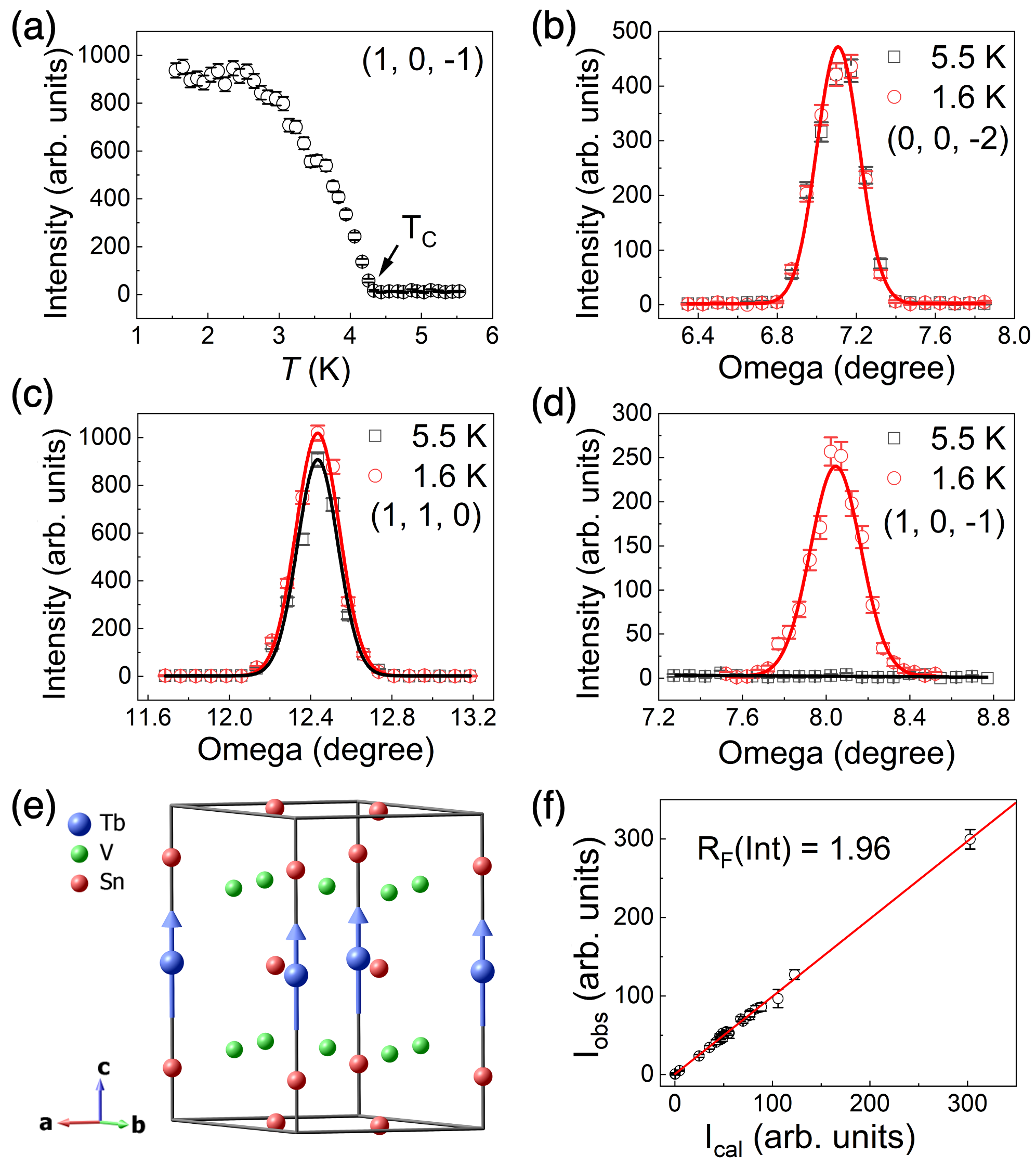}
\hspace{0cm}
\caption{\label{fig:Figs_TbV_neutron}
Single-crystal neutron diffraction results of TbV$_6$Sn$_6$. (a) Temperature dependence from 1.6 K to 5.5 K of (1, 0, -1). (b-d) Omega scans at 5.5 K and 1.6 K of (0, 0, -2), (1, 1, 0), and (1, 0, -1). (e) Magnetic structure of TbV$_6$Sn$_6$ with the magnetic propagation vector \textbf{k} = (0, 0, 0), and (f) fitted reflections as a function of observation and calculation.
}	
\end{figure}

The single-crystal neutron diffraction results of TbV$_6$Sn$_6$ are shown in Fig. \ref{fig:Figs_TbV_neutron}. As shown in Fig. \ref{fig:Figs_TbV_neutron}(a), it can be seen that the magnetic phase transition takes place at 4.3 K. The (1, 0, -1) reflection is chosen to do the temperature dependent measurement for the reason that the contribution from structural intensity here is almost zero and the change in magnetic intensity can be seen clearly (see Fig. \ref{fig:Figs_TbV_neutron}(d)). The intensity of the (0, 0, 2) reflection shows no difference between 5.5 K and 1.6 K, indicating that the magnetic moment is strictly aligned along the \textit{c} axis, which is consistent with the magnetization results (see Fig. \ref{fig:Figs_magnetic_properties}). A ferromagnetic order with the ordered magnetic moment aligned along the \textit{c} axis at the ground state is thus suggested. A total of 135 reflections are collected at both 5.5 K and 1.6 K for the refinement of the magnetic structure. Given the magnetic propagation vector \textbf{k} = (0, 0, 0) and a strong $c$-axis Ising anisotropy, the Irrep $\Gamma_{3}$ with the basis vector $\psi_{1} = (0, 0, 1)$ is used for magnetic structure refinement of TbV$_6$Sn$_6$. The ordered magnetic moment of Tb$^{3+}$ is determined at 9.4(2) $\mu_B$, which is very close to the effective moment of the free Tb$^{3+}$. The magnetic structure of TbV$_6$Sn$_6$ is shown in Fig. \ref{fig:Figs_TbV_neutron}(e), in which the magnetic moment of Tb$^{3+}$ is indeed aligned along the \textit{c} axis, thus further confirming the strong easy $c$-axis anisotropy as suggested previously \cite{pokharel_highly_2022, rosenberg_uniaxial_2022}.

\begin{figure}[htpb]
\centering
\includegraphics[width=8.5cm]{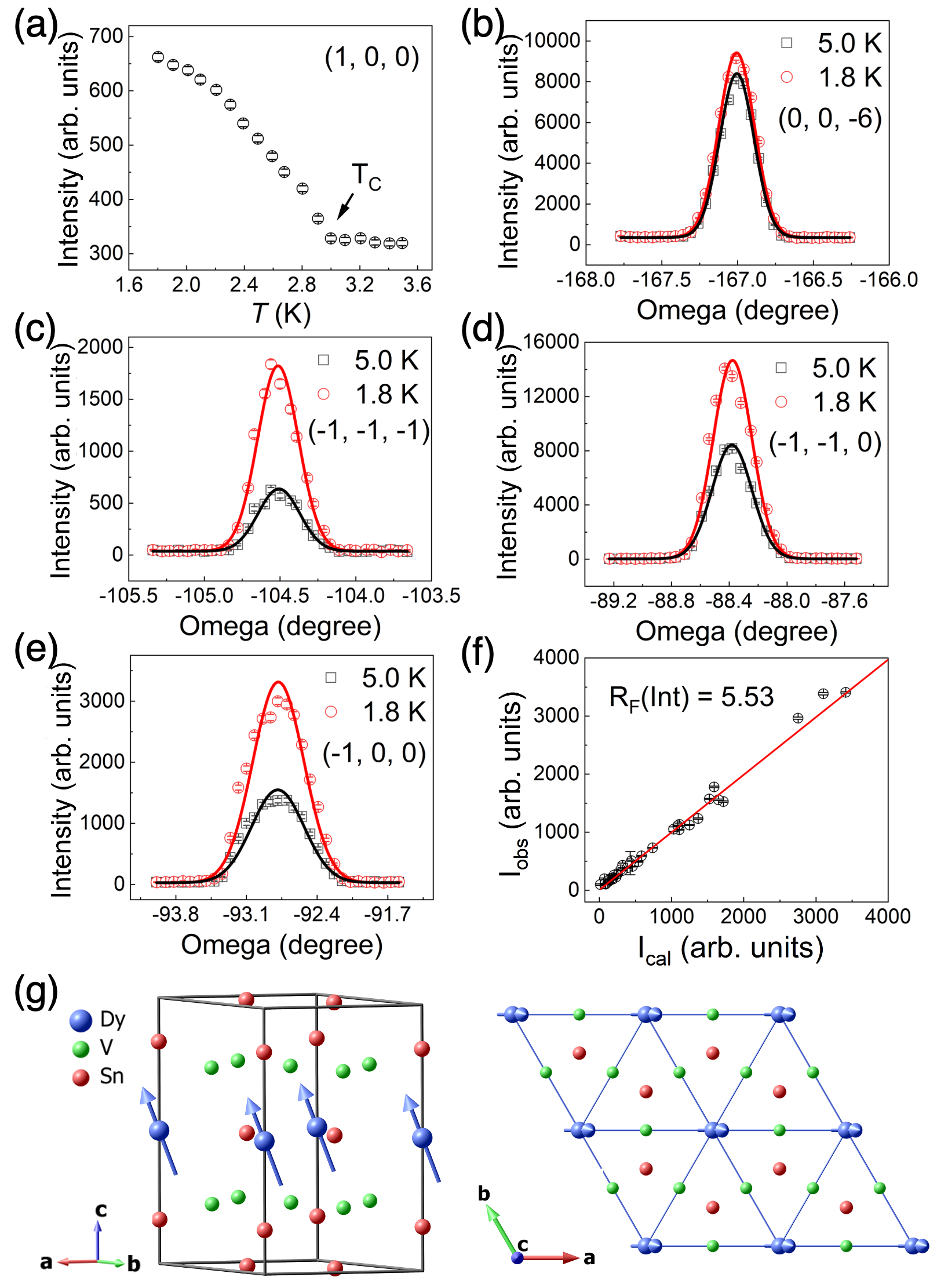}
\hspace{0cm}
\caption{\label{fig:Figs_DyV_neutron}
Single-crystal neutron diffraction results of DyV$_6$Sn$_6$. (a) Temperature dependence from 1.8 K to 3.5 K of (1, 0, 0). (b-e) Omega scans at 5.0 K and 1.8 K of (0, 0, -6), (-1, -1, -1), (-1, -1, 0), and (-1, 0, 0), respectively. (g) Magnetic structure of DyV$_6$Sn$_6$ with the propagation vector \textbf{k} = (0, 0, 0) in custom and top view, and (f) fitted reflections as a function of observation and calculation.
}	
\end{figure}

\subsection{Magnetic Structure of DyV$_6$Sn$_6$}

Fig. \ref{fig:Figs_DyV_neutron} presents the single-crystal neutron diffraction results of DyV$_6$Sn$_6$. The temperature dependence of the (1, 0, 0) reflection clearly reveals a magnetic phase transition at 3.0 K (see Fig. \ref{fig:Figs_DyV_neutron}(a)). A total of 245 reflections are measured to determine the magnetic structure at both 5.0 K and 1.8 K. Among these reflections, as shown in Fig. \ref{fig:Figs_DyV_neutron}(b-e), a difference in intensity between 5.0 K and 1.8 K for the (0, 0, -6) reflection can be seen, suggesting that a projection of the ordered magnetic moment in the \textit{ab} plane also exists. Similar intensity changes are also observed in the (-1, -1, -1), (-1, -1, 0), and (-1, 0, 0) reflections, which indicates the ordered magnetic moments of Dy$^{3+}$ at 1.8 K is neither simply aligned along the \textit{c} axis nor within the \textit{ab} plane, but tilted away from them. A weak magnetic anisotropy is also suggested from the magnetization measurements of DyV$_6$Sn$_6$ (see Fig. \ref{fig:Figs_magnetic_properties}(f)). Given the magnetic propagation vector \textbf{k} = (0, 0, 0) and the identified weak magnetic anisotropy, both the Irreps $\Gamma_{3}$ and $\Gamma_{9}$ including the basis vectors $\psi_{1} = (0, 0, 1)$, $\psi_{2} = (1, 0, 0)$ and $\psi_{3} = (1, 2, 0)$ are used in the refinement for DyV$_6$Sn$_6$, which is based on the measured 245 reflections after the neutron absorption correction. Furthermore, the magnetic structure refinement based on the symmetry-imposed and equally populated magnetic domains is also undertaken and the same result is obtained.  The solved magnetic structure reveals a magnetic moment of Dy$^{3+}$ as 6.6(2) $\mu_B$ ferromagnetically ordered inclined approximately 20${^\circ}$ off from the \textit{c} axis towards the [1, 0, 0] direction, as depicted in Fig. \ref{fig:Figs_DyV_neutron}(g).

\begin{figure}[htpb]
\centering
\includegraphics[width=8.5cm]{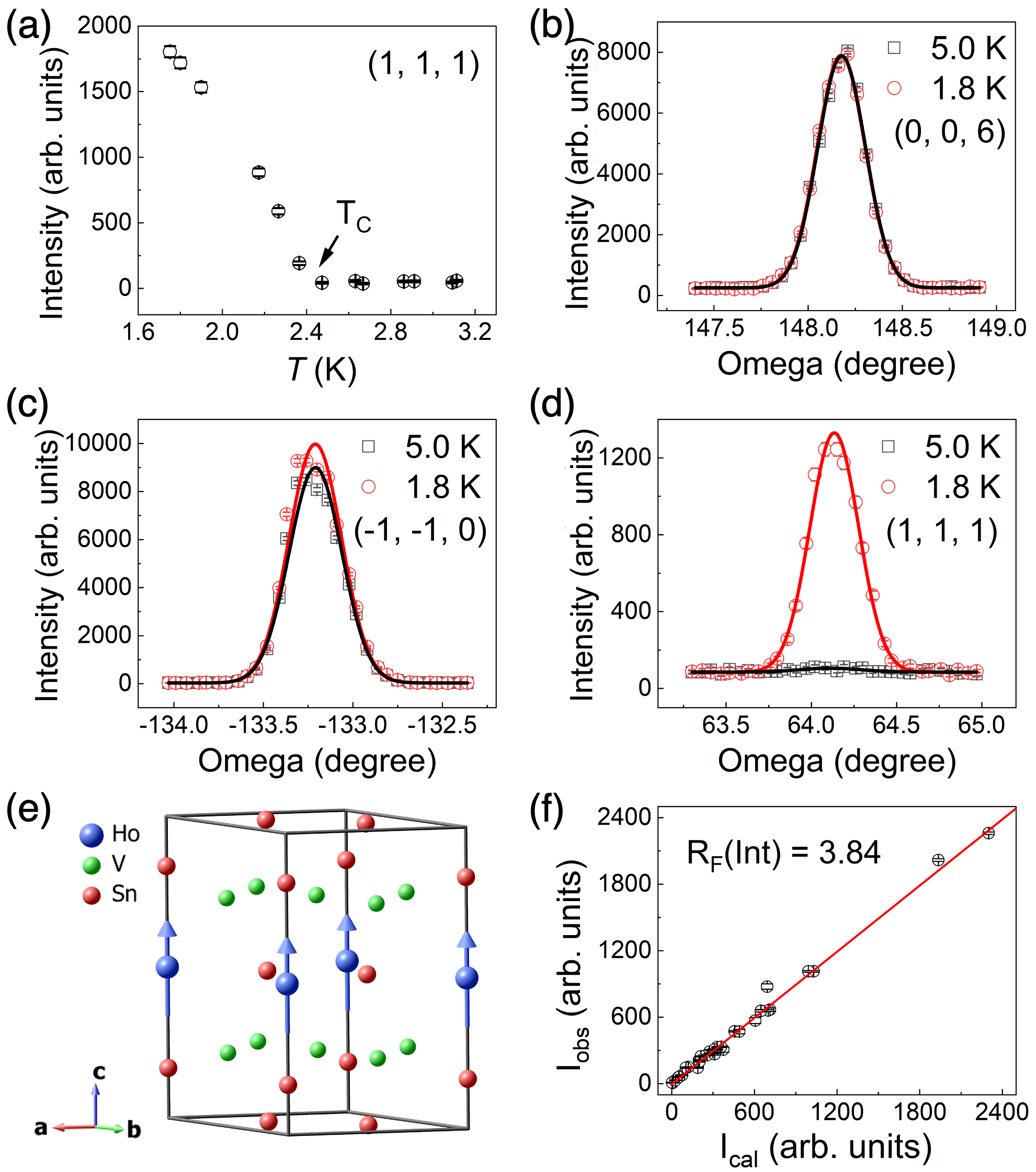}
\hspace{0cm}
\caption{\label{fig:Figs_HoV_neutron}
Single-crystal neutron diffraction results of HoV$_6$Sn$_6$. (a) Temperature dependence from 1.8 K to 3.1 K of (1, 1, 1). (b-d) Omega scans at 5.0 K and 1.8 K of the (0, 0, 6), (-1, -1, 0), and (1, 1, 1) reflections, respectively. (e) Magnetic structure of HoV$_6$Sn$_6$ with the magnetic propagation vector \textbf{k} = (0, 0, 0), and (f) fitted reflections as a function of observation and calculation.
}	
\end{figure}

\subsection{Magnetic Structure of HoV$_6$Sn$_6$}

The single-crystal neutron diffraction results for HoV$_6$Sn$_6$ are displayed in Fig. \ref{fig:Figs_HoV_neutron}. The (1, 1, 1) reflection is selected for the temperature dependence measurement. As shown in Fig. \ref{fig:Figs_HoV_neutron}(a), the onset of the magnetic phase transition can be identified at 2.4 K. For the (0, 0, 6) reflection, similar to the case in TbV$_6$Sn$_6$, no noticeable intensity variation between 1.8 K and 5.0 K can be seen, thus suggesting that the ordered magnetic moment is aligned in parallel to the \textit{c} axis. This finding is consistent with the magnetization results (see Fig. \ref{fig:Figs_magnetic_properties}(g)). Additionally, an increase in intensity is observed for the (-1, -1, 0) and (1, 1, 1) reflections at 1.8 K, which further supports that the magnetic moment of Ho$^{3+}$ is aligned along the \textit{c} axis. A total of 201 reflections were collected at both 5.0 K and 1.8 K. Based on the magnetic propagation vector \textbf{k} = (0, 0, 0) and a strong $c$-axis magnetic anisotropy, the Irrep $\Gamma_{3}$ with the basis vector $\psi_{1} = (0, 0, 1)$ is adopted for the magnetic structure refinement of HoV$_6$Sn$_6$. The final analysis yields an ordered magnetic moment of Ho$^{3+}$ as 6.4(2) $\mu_B$, along the \textit{c} axis. The magnetic structure is shown in Fig. \ref{fig:Figs_HoV_neutron}(e).

\subsection{Magnetic Structure of ErV$_6$Sn$_6$}

\begin{figure}[htpb]
\centering
\includegraphics[width=8.5cm]{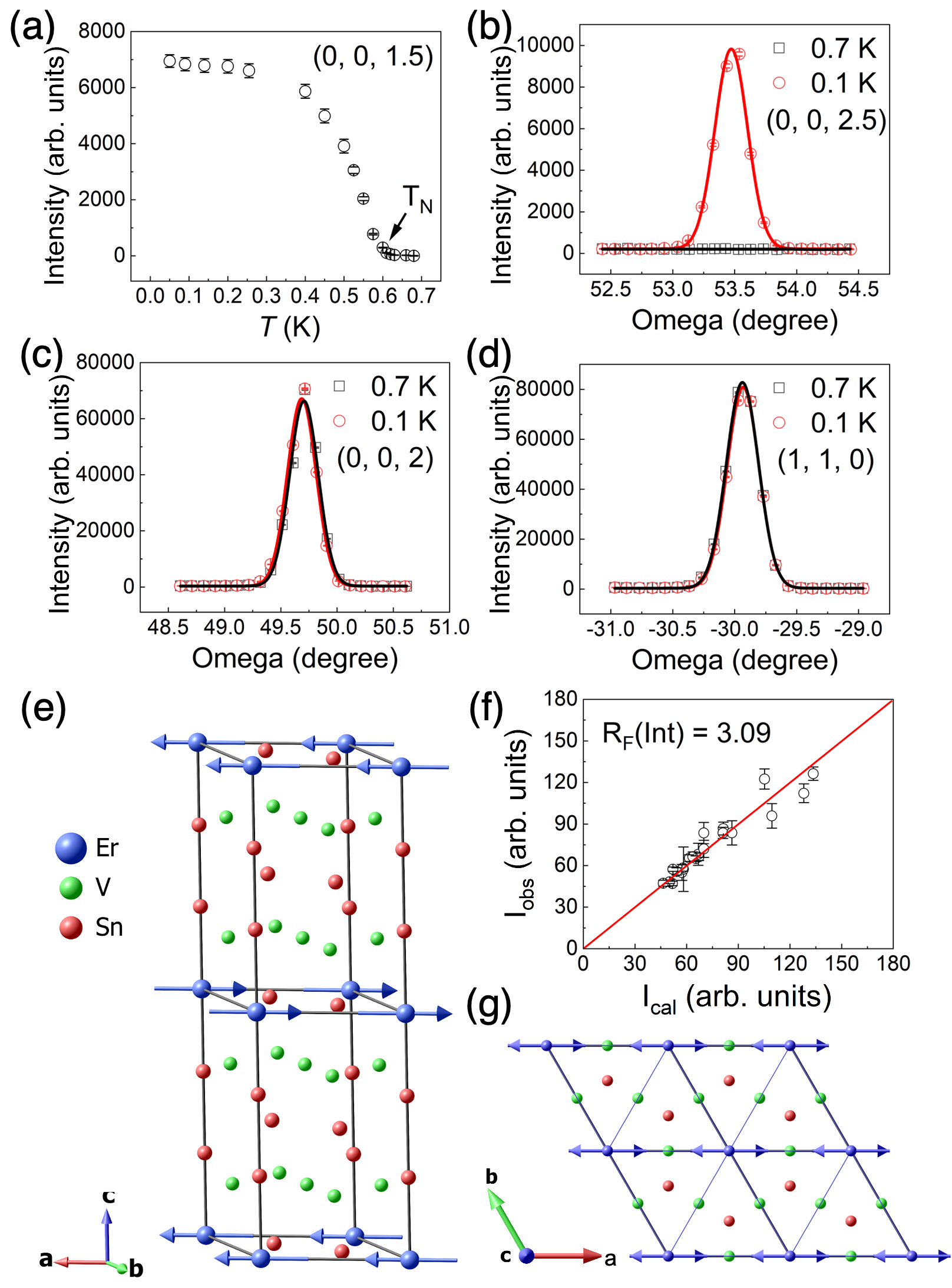}
\hspace{0cm}
\caption{\label{fig:Figs_ErV_neutron}
Single-crystal neutron diffraction results of ErV$_6$Sn$_6$. (a) Temperature dependence from 0.05 K to 0.7 K of (0, 0, 1.5). (b-d) Omega scans at 0.1 K and 0.7 K of the (0, 0, 2.5), (0, 0, 2), and (1, 1, 0) reflections, respectively. Magnetic structure of ErV$_6$Sn$_6$ with the magnetic propagation vector \textbf{k} = (0, 0, 0.5) in (e) custom and (g) top view, and (f) fitted reflections as a function of observation and calculation.
}	
\end{figure}

The single-crystal neutron diffraction results for ErV$_6$Sn$_6$ are shown in Fig. \ref{fig:Figs_ErV_neutron}. Magnetic reflections are found at the ($H$, $K$, $L {\pm} 0.5$) positions. The temperature dependence measurement at (0, 0, 1.5) reveals that the magnetic transition happens at 0.6 K (see Fig. \ref{fig:Figs_ErV_neutron}(a)). Omega scans conducted at 0.1 K and 0.7 K on the (0, 0, 2.5) reflection are depicted in Fig. \ref{fig:Figs_ErV_neutron}(b). At 0.1 K, a magnetic peak emerges at (0, 0, 2.5), indicating an antiferromagnetic structure with a magnetic propagation vector \textbf{k} = (0, 0, 0.5). The observed very weak intensity at the structural reflection (0, 0, 5) further rules out the possibility of second-order contamination originating from the structural reflections. Taking into account the significant magnetic anisotropy in the \textit{ab} plane from its magnetization properties, it is suggested that the magnetic moment of Er$^{3+}$ lies within the \textit{ab} plane. In total, 211 structural and 148 magnetic reflections are measured to determine the magnetic structure at 50 mK. The neutron absorption correction is also undertaken for all the measured reflections via Mag2Pol. Based on the magnetic propagation vector \textbf{k} = (0, 0, 0.5) and the observed strong easy $ab$-plane magnetic anisotropy, the Irrep $\Gamma_{9}$ with the basis vectors $\psi_{2} = (1, 0, 0)$ and $\psi_{3} = (1, 2, 0)$ are used in the refinement for ErV$_6$Sn$_6$. The custom and top view of the solved magnetic structure is shown in Fig. \ref{fig:Figs_ErV_neutron}(e) and (g). The final result reveals an ordered magnetic moment of Er$^{3+}$ as 6.1(3) $\mu_B$ in its ground state, and an A-type antiferromagnetic order within the \textit{ab} plane parallel to the \textit{a} direction.

\section{Discussions and outlook}

\begin{figure*}[htpb]
\centering
\includegraphics[width=16cm]{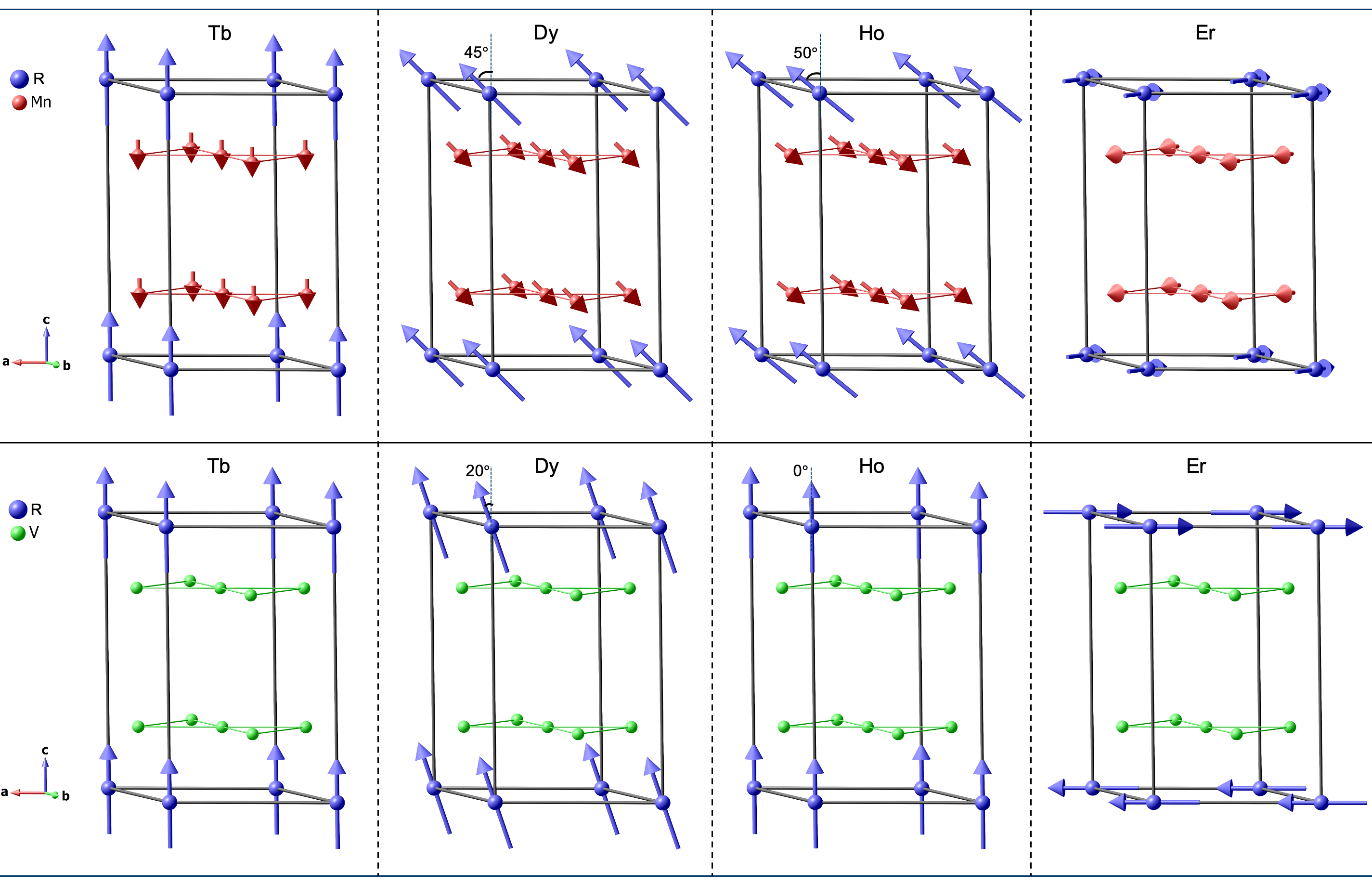}
\hspace{0cm}
\caption{\label{fig:Figs_magnetic_structures}
Comparison of the ground-state magnetic structures of \textit{R}Mn$_6$Sn$_6$ \cite{el_idrissi_magnetic_1991, malaman_magnetic_1999, mielke_iii_low-temperature_2022} and \textit{R}V$_6$Sn$_6$ (\textit{R} = Tb, Dy, Ho, Er).
}	
\end{figure*}

By investigating \textit{R}V$_6$Sn$_6$, we not only uncover the intriguing physical properties of a new series of topological kagome metals, as discussed in the previous sections, but also could gain new insights into the complex magnetic orders and interactions in another extensively investigated magnetic kagome metal series of \textit{R}Mn$_6$Sn$_6$. This is because \textit{R}V$_6$Sn$_6$ represents a simplified case where Mn is replaced by nonmagnetic V, thus switching off the magnetism from Mn as well as the magnetic interactions between \textit{R} and Mn. In this regard, YMn$_6$Sn$_6$ could serve as another simplified case for the understanding of the complex magnetism in \textit{R}Mn$_6$Sn$_6$, as the magnetism from \textit{R} and the corresponding \textit{R}-Mn interactions are also removed. 

Numerous neutron scattering studies on YMn$_6$Sn$_6$ \cite{venturini_incommensurate_1996, ghimire_competing_2020, dally_chiral_2021, wang_field-induced_2021-1} have revealed that under zero field it transforms initially to a commensurate AFM order with a propagation vector of (0, 0, 0.5) below 345 K, and, upon cooling to the base temperature, subsequently to a double-flat-spiral incommensurate magnetic order. This behavior is attributed to the strong intralayer Mn-Mn FM interaction due to a short distance ($\sim$2.8 \AA) between the neighboring Mn ions, and the competing interlayer magnetic interactions owing to the slight difference in distances between the nonequivalent Mn layers (i.e., [Mn-Sn2-Sn1-Sn2-Mn] and [Mn-\textit{R}Sn3-Mn] layers) \cite{rosenfeld_double-flat-spiral_2008, venturini_incommensurate_1996}.

In \textit{R}Mn$_6$Sn$_6$ (\textit{R} = Tb, Dy, Ho, Er), magnetic orders vary significantly due to different \textit{R}-Mn exchange interactions \cite{venturini_magnetic_1991, malaman_magnetic_1999, el_idrissi_magnetic_1991, kimura_high-field_2006, mielke_iii_low-temperature_2022, wang_real-space_2023, dhakal_anisotropically_2021, riberolles_new_2024}. For \textit{R} = Tb, Dy, and Ho, the \textit{R} and Mn moments exhibit FiM order below 423 K, 393 K, and 376 K, respectively, with a subsequent spin reorientation at 320 K (Tb), 277 K (Dy), and 195 K (Ho). In contrast, for ErMn$_6$Sn$_6$, an incommensurate magnetic spiral order appears below 345 K, which is then followed by an FiM order at 68 K. The high magnetic ordering temperatures in \textit{R}Mn$_6$Sn$_6$ are due to the strong intralayer Mn-Mn and interlayer \textit{R}-Mn interactions, while the spin reorientation transition is likely driven by the delicate balance between the magnetic anisotropy energy and the \textit{R}-Mn interaction. Thus, the \textit{R}-Mn interaction plays a crucial role in the \textit{R}Mn$_6$Sn$_6$ system with the distance of $\sim$3.6 \AA\ between the neighboring \textit{R} and Mn ions.

In \textit{R}V$_6$Sn$_6$, in which Mn is replaced by nonmagnetic V, the magnetic interactions become simpler, involving only intralayer and interlayer \textit{R}-\textit{R} interactions. The indirect Ruderman-Kittel-Kasuya-Yosida (RKKY) exchange interaction dominates due to the relatively long distances between \textit{R} ions (i.e., $\sim$5.5 \AA\ for intralayer and $\sim$9.0 \AA\ for interlayer \textit{R}-\textit{R} interactions). In contrast to the strong Mn-Mn and $R$-Mn direct exchange interactions in \textit{R}Mn$_6$Sn$_6$, the RKKY interaction is much weaker, so that the magnetic phase transition temperature is strongly suppressed in \textit{R}V$_6$Sn$_6$. The ground-state magnetic structures of \textit{R}Mn$_6$Sn$_6$ \cite{el_idrissi_magnetic_1991, malaman_magnetic_1999, mielke_iii_low-temperature_2022} and \textit{R}V$_6$Sn$_6$ (\textit{R} = Tb, Dy, Ho, Er) are summarized in Fig. \ref{fig:Figs_magnetic_structures}. A comparison of them could help to unravel the crucial role of the $4f$-$3d$ \textit{R}-Mn interactions played in the tuning of the magnetic properties of the \textit{R}Mn$_6$Sn$_6$ system. The distinct magnetic anisotropy of various \textit{R} ions shows evidently in \textit{R}V$_6$Sn$_6$: the ordered magnetic moment is aligned along the $c$ axis for both TbV$_6$Sn$_6$ and HoV$_6$Sn$_6$, slightly tilted away from the $c$ axis for DyV$_6$Sn$_6$, and aligned within the $ab$ plane for ErV$_6$Sn$_6$, respectively. The ordering temperatures for \textit{R}V$_6$Sn$_6$ are 4.3 K, 3.0 K, 2.4 K, and 0.6 K for the Tb, Dy, Ho, and Er compounds, respectively (see Tab. \ref{tab:neutron}). These temperatures follow the same trend as the spin-reorientation transition temperatures in \textit{R}Mn$_6$Sn$_6$ that decrease with increasing atomic number of the rare-earth elements, indicating a weaker \textit{R}-Mn interaction for rare-earth elements with higher atomic numbers. This trend also suggests that the \textit{R}-Mn interaction is significantly influenced by the intrinsic single-ion properties of \textit{R}$^{3+}$, which strongly depend on crystalline electric field (CEF) effects. Additionally, the large difference in magnetic transition temperature between \textit{R}V$_6$Sn$_6$ and \textit{R}Mn$_6$Sn$_6$ further indicates a robust \textit{R}-Mn interaction.

There are numerous possible emergent quantum phenomena yet to be discovered in this system, particularly concerning the interplay between topology and magnetism. Unlike the \textit{R}Mn$_6$Sn$_6$ system, where the kagome layers are magnetic, the \textit{R}V$_6$Sn$_6$ system features a non-magnetic topological kagome layer and a magnetic triangular layer of \textit{R}, thus providing a fresh platform for understanding such an interplay. For instance, when an out-of-plane magnetization is introduced to the kagome lattice in TbMn$_6$Sn$_6$, it lifts the spin degeneracy, thus transforming the Z$_2$ topological gap into a quantum-limit Chern gap, and resulting in the kagome lattice hosting exotic chiral edge states \cite{yin_quantum-limit_2020,yin_topological_2022}. However, in the \textit{R}Mn$_6$Sn$_6$ system, the kagome layers are inherently magnetic, which breaks time-reversal symmetry and may significantly alter the electronic structure of the kagome lattice. This condition does not occur in the \textit{R}V$_6$Sn$_6$ system, making it a perfect platform to study the interplay between magnetism and topology. The magnetic nature of \textit{R} in the \textit{R}V$_6$Sn$_6$ system can then be used as an ideal tuning parameter to explore how magnetism influences the topological features arising from the kagome geometry, offering insights that are unattainable in magnetic systems like \textit{R}Mn$_6$Sn$_6$. For instance, it would be very interesting to search for possible exotic quantum states like Chern insulator or Weyl semimetal at its ground state at very low temperatures in TbV$_6$Sn$_6$, owing to the ferromagnetically ordered large moment of Tb$^{3+}$ along the \textit{c} axis and the nonmagnetic V kagome layers. HoV$_6$Sn$_6$ may exhibit a similar but relatively weaker effect. Additionally, the spin Berry curvature \cite{di_sante_flat_2023}, that is related to the bilayer kagome lattice, may also be systematically explored in the magnetically ordered states in \textit{R}V$_6$Sn$_6$.

Another interesting aspect to discuss is the magnetic frustration in \textit{R}V$_6$Sn$_6$ due to the presence of an ideal triangular lattice of the \textit{R} site. However, the dominating FM intralayer \textit{R}-\textit{R} interaction does not provide a strong basis for frustration, especially in the Ising-type ordered compound TbV$_6$Sn$_6$, with a magnetic moment of 9.4(2) $\mu_B$ ordered along the \textit{c} axis. For DyV$_6$Sn$_6$ and HoV$_6$Sn$_6$, their magnetic moments at 1.8 K are still notably smaller than those of the free ions. In ErV$_6$Sn$_6$, the ordered magnetic moment is 6.1(3) $\mu_B$ at 50 mK, which is also much smaller than its free ion moment. This thus hints a likely strong influence of the single-ion CEF effects and possible presence of persistent spin fluctuations in the ground state. Clearly, further studies of CEF excitations and spin dynamics would be of high interests.

\section{Conclusions}

In this work, we report mainly a comprehensive single-crystal neutron diffraction investigation of the ground-state magnetic structures of the recently discovered V-based topological kagome metals \textit{R}V$_6$Sn$_6$ (\textit{R} = Tb, Dy, Ho, Er). We also systematically investigated the crystal structure, magnetic properties and specific heat capacity (down to 50 mK) of the high-quality single-crystal samples of this kagome metal series, that were successfully grown by the flux method, via a wide range of in-house characterization techniques. These in-house investigations indicate the presence of long-range magnetic order at low temperatures. Our single-crystal neutron diffraction measurements further confirm that the long-range magnetic order indeed takes place at 4.3 K (Tb), 3.0 K (Dy), 2.4 K (Ho), and 0.6 K (Er), respectively, and no further magnetic phase transitions were found at lower temperatures down to 50 mK according to the heat capacity results. The ground-state magnetic structures of all the studied compounds are comprehensively determined via the magnetic crystallography approaches.  It can be revealed that \textit{R}V$_6$Sn$_6$ (\textit{R} = Tb, Dy, Ho) transform to a collinear ferromagnetic order below the magnetic phase transition temperature, with the ordered magnetic moment aligned along the \textit{c} axis for \textit{R} = Tb, Ho, while approximately 20${^\circ}$ tilted off from the \textit{c} axis for \textit{R} = Dy. Due to an apparent antiferromagnetic interlayer coupling between triangular lattice Er layers, ErV$_6$Sn$_6$ shows an A-type antiferromagnetic structure with a magnetic propagation vector \textbf{k} = (0, 0, 0.5), and with the ordered moment aligned in the \textit{ab} plane. The ordered magnetic moment are determined as 9.4(2) ${\mu_B}$, 6.6(2) ${\mu_B}$, 6.4(2) ${\mu_B}$, and 6.1(2) ${\mu_B}$ for \textit{R} = Tb, Dy, Ho, and Er, respectively. We also compared and discussed the low-temperature magnetic structures in both  \textit{R}V$_6$Sn$_6$ and \textit{R}Mn$_6$Sn$_6$ kagome metal series. This allows to gain new insights into the complex magnetic interactions, single-ion magnetic anisotropy and spin dynamics in these compounds. With the determined ground-state magnetic structures in these four compounds, further investigations on the possible interplay between magnetism and topologically non-trivial electron band structures in the magnetically ordered phase regime can be expected for this fascinating topological kagome metal series of \textit{R}V$_6$Sn$_6$.

\section*{Acknowledgments}

This work is based on a series of single-crystal neutron diffraction experiments performed at Sika  (ANSTO, Sydney), Zebra (SINQ, PSI, Villigen), and D23 (ILL, Grenoble) neutron instruments. The single-crystal growth, EDX and XRD measurements were performed at JCNS-MLZ, Garching. The heat capacity was measured with a PPMS device and a dilution refrigerator system at NCKU. We would like to thank the late Marie-Sousai Appavou for his great help on the EDX measurements that made this work possible. We would also like to acknowledge T. Schrader for the assistance on the XRD measurements, and V. Ray, S. Nandi and O. Petracic for their assistances in the preliminary characterization of our single-crystal samples, and N. Qureshi for valuable discussions on Mag2Pol. Y.Z. acknowledges the scholarship funding from the Chinese Scholarship Council. S.H. acknowledges the postdoctoral funding from the Palestinian-German Science Bridge (PGSB) program. S.D. acknowledges the postdoctoral funding from the European Union's Horizon 2020 research and innovation programme under the Marie Skłodowska-Curie grant agreement No 101034266. C.H.H. acknowledges the funding from the Postdoctoral Research Abroad Program from National Science and Technology Council in Taiwan. \\

\bibliography{YZ_ref.bib}

\end{document}